# In plane quantification of *in vivo* muscle elastic anisotropy factor by steered ultrasound pushing beams

Ha-Hien-Phuong Ngo[1], Ricardo Andrade[2], Javier Brum[3], Nicolas Benech[3], Simon Chatelin[4], Aude Loumeaud[4], Thomas Frappart[5], Christophe Fraschini[5], Antoine Nordez[2,6], Jean-Luc Gennisson[1]

[1]Laboratoire d'imagerie médicale multimodale, BioMaps, Université Paris Saclay, CEA, CNRS, Inserm, Orsay, France,

[2]Nantes Université, Mouvement - Interactions - Performance, MIP, UR 4334, F-44000 Nantes, France

[3]Laboratorio de Acústica Ultrasonora, Instituto de Física, Facultad de Ciencias, Universidad de la República, Montevideo, Uruguay,

[4]ICube, CNRS UMR 7357, University of Strasbourg, Strasbourg, France

[5]Hologic - Supersonic Imagine, Aix-en-Provence, France,

[6]Institut Universitaire de France (IUF), Paris, France

Corresponding author: jean-luc.gennisson@universite-paris-saclay.fr

Keywords: elastography, anisotropy, steered pushing beam, shear waves, muscles

**Abstract (300 words)**

Skeletal muscles are organized into distinct layers and exhibit anisotropic characteristics across various scales. Assessing the arrangement of skeletal muscles may provide valuable biomarkers for diagnosing muscle-related pathologies and evaluating the efficacy of clinical interventions. In this study, we propose a novel ultrafast ultrasound sequence constituted of steered pushing beams was proposed for ultrasound elastography applications in transverse isotropic muscle. Based on the propagation of the shear wave vertical mode, it is possible to fit the experimental results to retrieve in the same imaging plane, the shear modulus parallel to fibers as well as the elastic anisotropy factor (ratio of Young's moduli times the shear modulus perpendicular to fibers). The technique was demonstrated *in vitro* in phantoms and *ex vivo* in fusiform beef muscles. At last, the technique was applied *in vivo* on fusiform muscles (*biceps braachi*) and mono-penate muscles (*gastrocnemius medialis*) during stretching and contraction. This novel sequence provides access to new structural and mechanical biomarkers of muscle tissue, including the elastic anisotropy factor, within the same imaging plane. Additionally, it enables the investigation of multiples parameters during muscle active and passive length changes.

**Introduction**

Shear wave elastography is an ultrasound technique that enables the quantification of shear wave speed in soft tissues, which is directly linked to the tissue stiffness (Royer and Dieulesaint, 1999). Many techniques have been developed to generate shear waves in tissues with different devices (Catheline et al. 1999, Muthupillai *et al.*, 1995, Parker *et al.*, 1990, Grasland-Mongrain *et al.*, 2013) and to image their propagation with ultrasound (Fatemi and Greenleaf, 1998, Sandrin *et al.*, 2002, Urban *et al.*, 2011, Song *et al.*, 2012). In 1995, Sarvazyan *et al.* proposed to use the acoustic radiation force, also called pushing beam, as a shear wave source (Sarvazyan et al., 1995) allowing to control the shear wave source directly within tissues. Subsequently, several methods were developed based on this principle (Nigthingale *et al.*, 2003, Bercoff *et al.*, 2004). However, they primarily assume, as a first approximation, that tissues are purely elastic, locally homogeneous, and isotropic. While this assumption proves advantageous for various organs such as the liver or breast (Sandrin *et al.*, 2003, Berg *et al.*, 2012), it is important to note that most human organs, including skeletal muscles, exhibit inherent anisotropic characteristics.

Actually, muscles consist of multiple aligned fibers primarily oriented in one direction, known as the main direction, with random distribution in the transverse plane. This corresponds to the definition of a transverse isotropic (TI) medium where two elastic shear moduli can be defined: $\mu_\perp$ and $\mu_{//}$, shear modulus perpendicular and shear modulus parallel to fibers respectively (Royer and Dieulesaint, 1999). To quantify both coefficients, two potential solutions can be considered. The first solution involves moving the transducer array to acquire shear wave speed in different directions (Gennisson *et al.*, 2010, Royer *et al.*, 2011). While this approach is relatively straightforward in parallel muscles (fusiform muscles), it requires multiple data acquisitions (at least two acquisition, parallel and perpendicular to fibers), which can be challenging in practical scenarios involving clinical examinations and biomechanical investigations of muscle behavior during active or passive muscle length changes. In addition, it can be used only in fusiform muscles (i.e., fibers oriented in the muscle shortening direction), while the large majority of muscles are pennated.

More generally, all elastography measurements performed in a pennated muscle provide a shear modulus along the probe direction that does not fit to the fiber direction, making the interpretation quite complex when the fiber direction is changed due to muscle lengthening, contraction, or even due to an intervention and in a pathologic muscle.

Alternatively, a second approach involves utilizing a three-dimensional (3D) acquisition device (Wang et al., 2013, Gennisson *et al.*, 2015, Correia *et al.*, 2018, Knight *et al.*, 2022) to obtain elastography data in volume, enabling the exploration of multiple directions from a single dataset. However, the corresponding ultrasound devices are expensive and primarily developed for research purposes, making them currently unavailable in clinical settings.

In this study, we propose a novel two-dimensional (2D) ultrasound sequence, implemented in a clinical device, to investigate and quantify two elastic coefficients, in the imaging plane: the shear modulus parallel to the fibers ($\mu_{//}$) and an elastic anisotropy factor of muscles (the ratio of both Young' moduli times the shear modulus perpendicular to fibers). A steered ultrasound pushing beam (SPB) method was developed to generate shear waves with a certain propagation angle. The resulting shear wave propagation was captured using an ultrafast ultrasound acquisition sequence, which was beamformed along the same angle as the SPB. This paper is organized as follow: In the first part of our study presents the novel sequence proposed and the theoretical framework defining the propagation of shear wave with different propagation modes within a transverse isotropic medium. In the second part, we present the experimental setups and the novel ultrasound sequence. The validation of the proposed SPB technique was demonstrated *in vitro* in phantoms before being used to quantify the elastic anisotropy factor *ex vivo* and *in vivo* skeletal muscles. Finally, the elastic anisotropy factor was assessed in three healthy volunteers for two muscle types (fusiform and pennated) during active and passive muscle length changes.

**Theory**

*Shear wave propagation in transverse isotropic medium*

Before explaining and presenting the proposed SPB technique, a reminder of the theory of shear wave propagation in transverse isotropic media must be made. A TI elastic material is characterized by 5 independent elasticity constants in the elasticity tensor involved in the generalized Hooke's law relating the stress *σ* and the strain *ε* tensors, which can be expressed using Voigt's notation as:

$$\begin{bmatrix} \sigma_1 \\ \sigma_2 \\ \sigma_3 \\ \sigma_4 \\ \sigma_5 \\ \sigma_6 \end{bmatrix} = \begin{bmatrix} C_{11} & C_{12} & C_{13} & 0 & 0 & 0 \\ C_{12} & C_{11} & C_{13} & 0 & 0 & 0 \\ C_{13} & C_{13} & C_{33} & 0 & 0 & 0 \\ 0 & 0 & 0 & C_{44} & 0 & 0 \\ 0 & 0 & 0 & 0 & C_{44} & 0 \\ 0 & 0 & 0 & 0 & 0 & C_{66} \end{bmatrix} \begin{bmatrix} \varepsilon_1 \\ \varepsilon_2 \\ \varepsilon_3 \\ \varepsilon_4 \\ \varepsilon_5 \\ \varepsilon_6 \end{bmatrix} \quad (1)$$

with $C_{66} = (C_{11} - C_{12})/2$.

In a TI material elastic wave propagation is governed by a set of linear equations known as "Christoffel equations" which relate wave speed, propagation direction, polarization, and mechanical properties (Royer and Dieulesaint, 1999). Under plane wave decomposition of the form:

$$u_i(\vec{r}, t) = u_o \exp[i(\vec{k}.\vec{r} - \omega t)] \quad (2)$$

where $u_o$ is the wave amplitude, $\vec{k}$ stands for the wave vector, $\omega$ is the frequency and $u_i$ ($i$=1, 2, 3) denotes the spatial components of the displacement field (Figure 1a), one can obtain an eigenvalue equation for the matrix $\Gamma_{ik} = C_{ijkl} n_j n_l$,

$$|\Gamma_{ik} - \rho v^2 \delta_{ik}| = 0 \qquad (3)$$

With $\rho$ being the material density assumed to be constant (~1100 kg.m$^{-3}$) and phase velocity $v = \frac{\omega}{k}$. The tensor Γ, also known as Christoffel tensor, can be written for a TI material as

$$\Gamma = \begin{pmatrix} C_{11}n_1^2 + C_{66}n_2^2 + C_{44}n_3^2 & (C_{11} + C_{66})n_2 n_1 & (C_{44} + C_{13})n_1 n_3 \\ (C_{11} + C_{66})n_2 n_1 & C_{66}n_1^2 + C_{11}n_2^2 + C_{44}n_3^2 & (C_{44} + C_{13})n_2 n_3 \\ (C_{44} + C_{13})n_1 n_3 & (C_{44} + C_{13})n_2 n_3 & C_{44}n_1^2 + C_{44}n_2^2 + C_{33}n_3^2 \end{pmatrix} \qquad (4)$$

where $\vec{n}(n_1, n_2, n_3)$ is the propagation vector, *i.e.* $\vec{k} = k\vec{n}$. From this expression of the Christoffel tensor, 3 cases can be expressed (Royer and Dieulesaint, 1999, Ngo *et al.*, 2022): i) $\vec{n} = (1,0,0)$; ii) $\vec{n} = (0, \sin(\varphi), \cos(\varphi))$; iii) $\vec{n} = (\sin(\theta), 0, \cos(\theta))$.

*Shear vertical and shear horizontal modes*

In the present paper we focused on the particular case where the shear wave propagation vector lies within the ($x_1$, $x_3$)-plane (axis defined in figure 1), which can be written as $\vec{n} = (\sin(\theta), 0, \cos(\theta))$, $\vartheta$ being the angle between the propagation vector and the fiber orientation. In this example the eigenvalues are given by:

$$\rho v_{1,3}^2 = \frac{1}{2}\left[\Gamma_{11} + \Gamma_{33} \mp \sqrt{(\Gamma_{33} - \Gamma_{11})^2 + 4\Gamma_{13}^2}\right] \quad (5)$$

(where the minus sign corresponds to the first eigenvalue) and:

$$\rho v_2^2 = \Gamma_{22} = C_{66}\sin^2(\theta) + C_{44}\cos^2(\theta) = \mu_\perp \sin^2(\theta) + \mu_\parallel \cos^2(\theta) \quad (6)$$

The second eigenvalue corresponds to a pure shear wave, *i.e.* with eigenvector parallel to ($x_2$), also known a shear horizontal (SH) mode (Rouze *et al.*, 2020, Caenen *et al.*, 2020). Because for this mode the displacement is perpendicular to the imaging plane (*i.e.* ($x_1$, $x_3$)-plane), it will have only a negligible impact on the displacement field measured by the SPB configuration. The first and third eigenvalues correspond to a quasi-shear and a quasi-longitudinal wave polarized in the ($x_1$, $x_3$)-plane. This quasi-shear wave mainly dominates the displacement field generated by the steered push configuration. In the incompressibility limit, the phase velocity for this quasi-shear mode reduces to (Rouze *et al.*, 2013):

$$\rho v_1^2 = C_{44} + \left(\frac{C_{33}C_{11}-C_{13}^2}{4(C_{11}-C_{66})} - C_{44}\right)\sin^2(2\varphi) = \mu_L + \left(\frac{E_\parallel}{E_\perp}\mu_\perp - \mu_\parallel\right)\sin^2(2\varphi) \quad (7)$$

and its polarization is purely transverse, therefore, this type of mode is also known as shear vertical (SV) mode (Rouze *et al.*, 2020, Caenen *et al.*, 2020). Equations (6) and (7) can also be expressed in terms of $\mu_{//,\perp}$ and $E_{//,\perp}$ which refer to the longitudinal (//) and transverse (⊥) shear and Young's moduli, respectively. Hence, in the incompressibility limit, shear wave propagation in TI media only depends on three independent parameters: $\mu_{//}$, $\mu_\perp$ and the anisotropy ratio of the Young's moduli $\chi_E = E_{//}/E_\perp$. These three parameters are required to fully characterize the mechanical behavior of incompressible TI tissue and can be extracted from the SH and SV wave propagation (Thomsen, 1986).

*Phase velocity vs. Group velocity in steered push beam configuration*

Equations (6) relates the phase velocity of the SH mode with the mechanical parameters of the TI medium. In the following manuscript only, the angles θ=0° and θ=90° were measured, corresponding to the position of the transducer array parallel to fibers ($x_1$, $x_3$)-plane, Fig. 1a) and the perpendicular to fibers ($x_1$, $x_2$)-plane respectively. In that specific two cases the phase velocity is equal to the group velocity (Royer and Dieulesaint, 1999).

Equation (7) relates the phase velocity of the SV mode with the mechanical parameters of the TI medium. At this point it is important to mention the distinction between phase (*v*) and group ($V_g$) velocities in TI tissues. While in a non-dispersive isotropic tissue phase and group velocity are equal, in a TI tissue phase and group velocities magnitudes and directions in the same point are generally not (Thomsen, 1986, Royer and Dieulesaint, 1999). The phase velocity *v(α)* is also called the wavefront velocity since it measures the velocity of advance of the wavefront along $\vec{k}(\alpha)$. In general, for non-spherical wavefronts, the angle $\alpha$ is different from the ray angle $\varphi$ from the source point to the wavefront. Thus, the relation between ray angle and ray velocity, also called group angle and group velocity, respectively, is expressed as (Thomsen 1986):

$$\tan(\varphi(\alpha)) = \frac{\tan(\alpha)+dv/d\alpha}{1-\frac{\tan(\alpha)}{v}dv/d\alpha} \quad (8)$$

and $\rho V_g^2(\varphi(\alpha)) = \rho\left(v^2(\alpha) + \left(\frac{dv}{d\alpha}\right)^2\right)$ (9)

Thus, by combining equations (7) and (9) the following expression was found for the group velocity of the SV modes:

$$V_g^2(\varphi(\alpha)) = v^2(\alpha) + \mu_{//} \frac{d^2 + sin^2(4\alpha)}{1 + d.sin^2(2\alpha)} = \frac{1}{\rho}[\mu_L + (\chi_E \mu_\perp - \mu_\parallel)\sin^2(2\alpha)] + \mu_{//} \frac{d^2 + sin^2(4\alpha)}{1 + d.sin^2(2\alpha)} \quad (10)$$

with $d = \frac{\chi_E \mu_\perp}{\mu_\parallel} - 1$.

**Materials and methods**

*Ultrasound acquisitions*

An ultrafast ultrasound device (Mach30, Hologic - Supersonic Imagine, Aix-en-Provence, France) was used to drive a linear array transducer (SL18-5, 7.5 MHz, 256 elements). Two different ultrasound sequences were used to investigate horizontal-polarized (SH) and vertical-polarized (SV) modes of propagating shear waves, a commercial and a research sequence respectively. In both sequences shear waves were generated by focusing ultrasound onto tissues, which creates an acoustic radiation force (Sarvazyan *et al.*, 1995), commonly known as a pushing beam. This pushing beam acts as a shear wave source. Shear wave propagation was captured using an ultrafast ultrasound imaging sequence consisting of multiple compounded ultrasound plane waves (Montaldo *et al.*, 2009). The shear wave speed was computed by applying a time-of-flight algorithm (Loupas *et al.*, 1995, Deffieux *et al.*, 2012) to the recorded shear wave propagation movie (Bercoff *et al.*, 2004). For the commercial sequence the whole process is implemented in the ultrafast ultrasound device and the final shear modulus is obtained directly on screen. For the research sequence, the shear modulus map was obtained offline after downloading the raw data and processing under Matlab® (2022a, Mathworks, Natick, MA, USA) with a custom-made algorithm developed as for the commercial sequence from Loupas and Deffieux (Loupas *et al.*, 1995, Deffieux *et al.*, 2012).

The commercial sequence, implemented in the Mach30 device, was used to capture the propagation of the SH mode and the final shear modulus was obtained directly. In that configuration, the pushing beam was always perpendicular to the fibers and the propagation of the shear wave was measured parallel and perpendicular to the fibers (Gennisson *et al.*, 2011) (Figure 1a). In other words, in that case, the polarization of shear wave is always perpendicular to the fiber's axis (Royer *et al.*, 2011). The shear moduli $\mu_{//}$ parallel and $\mu_\perp$ perpendicular to the fibers' orientation, directly calculated by the commercial device from the shear wave velocity $V_S(\vartheta)$, were retrieved by using equation 5, where $\theta$ is the angle of rotation of the probe regarding the fiber's axis and $\rho$ the tissue density. The research sequence, referred to as the steered ultrasound pushing beam (SPB) sequence in this manuscript, was implemented in the ultrafast ultrasound device to capture the propagation of the SV mode. In this configuration, the ultrasound probe was positioned parallel to the axis of the fibers and remained fixed throughout the scans without any rotation (Figure 1b). The pushing beam as well as the beamforming that allows to reconstruct images were oriented with an angle $\varphi$ ranging from -20° to 20° by 5° steps. The shear wave source was constituted of 3 pushing beams of 150 µs spaced each other by 3 mm from 9 mm deep to deeper with a f-number of 1.75 for each. Briefly, for an angle $\varphi$, the pushing beam was created by small apertures with an oriented delay law of angle $\varphi$ (Figure 2a). Then, the ultrafast ultrasound sequence made of 5 tilted plane waves was oriented with the same angle $\varphi$ (from $\varphi$-4° to $\varphi$+4° by 2° step) with a framerate of 4300 Hz to catch the propagation of the shear waves. Mechanical parameters parallel and perpendicular shear moduli as well as the elastic anisotropy factor were retrieved by fitting the experimental data with equation 9. In the case of pinnate muscle, the transducer array is aligned with the main axis of fibers ($x_3$ axis) and the ultrasound sequence used is the SPB sequence (Figure 1c).

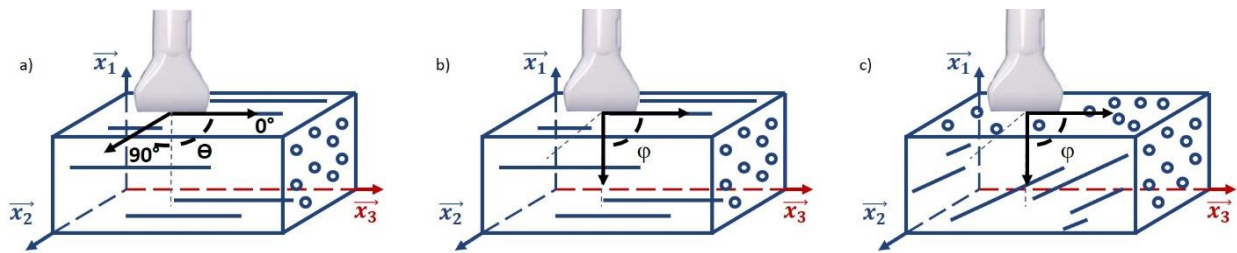

*Figure 1: a) Positioning of the ultrasound transducer array regarding the main axis of fibers to quantify the SH mode. The transducer can be rotated around $x_1$ axis by 90° to be aligned with axis $x_2$ or $x_3$ to quantify the shear modulus perpendicular or parallel to fibers respectively. b) Positioning of the transducer array parallel to the main fiber axis to quantify SV mode. In that configuration, the imaging plane is steered electronically of an angle $\varphi$ (from -20° to 20°) and the transducer remains stationary without any movement. c) SPB sequence used in the case of pinnate muscle when the main axis of muscle fiber is aligned with the transducer array along $x_3$ axis.*

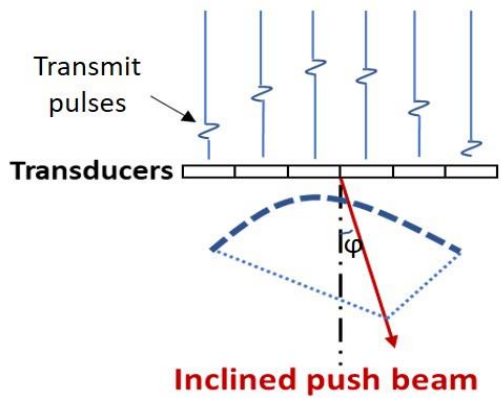
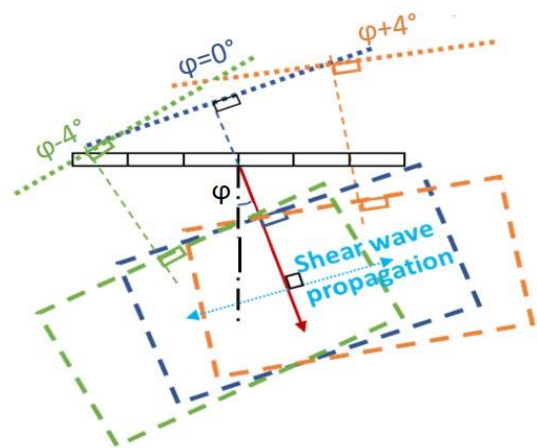

*Figure 2: Chronogram of the SPB sequence. a) A steered pushing beam is electronically programmed in the ultrafast ultrasound device to generate a steered acoustic radiation force that act as a shear wave source. b) ultrafast acquisition of the propagating shear wave with different tilted ultrafast ultrasound plane wave tilted around the axis of the pushing beam. In comparison, the ultrasound sequence implemented in the ultrafast ultrasound device in commercial mode is the same than SPB sequence with an angle $\varphi$ = 0°.*

Experimental results of shear wave group velocity were fit by using equation 10 and the fit function from Matlab with 95% confidence bounds. This fit has two variables which are $a = \mu_{//SPB}$ and $d$. The error of $d(\Delta d)$ was also given by the fitting process.

To calculate $b = (d+1).a = \chi_{EUS}.\mu_{\perp SPB}$ in order to deduce $\chi_{EUS} = \frac{b}{\mu_{\perp SSI}}$, the error of $b$ was calculated as :

$$\Delta b = \Delta\big((d+1).a\big) = \Delta d.a + \Delta a.d \quad (11)$$

Then the error of $\chi_{EUS}$ was calculated as follow:

$$\Delta\chi_{EUS} = \frac{|\Delta b.\mu_{\perp SSI} - b.\Delta\mu_{\perp SSI}|}{\mu_{\perp SSI}^2} \quad (12)$$

*In vitro and ex vivo experiments with the SPB sequence*

The SPB sequence was tested *in vitro* in an isotropic Polyvinyl Alcohol (PVA) phantom, that followed 4 freezing-thawing cycles and made of 10% PVA (molecular weight 89 000–98 000, 99+% hydrolyzed, Sigma-Aldrich, Saint-Louis, MO, USA) mixed with 2% cellulose particles that act as scatterers (20 µm in diameter, S3504 Sigmacell, Sigma-Aldrich, Saint-Louis, MO, U.S.A.) (Fromageau *et al.*, 2007; Chatelin et al., 2014) (Figure 3(a)). First *ex vivo* experiments were performed on beef muscle bought at the butcher shop to test the SPB sequence. This muscle sample was specifically chosen due to its fibrous structure, which exhibits a high degree of anisotropy with clearly visible uniaxial alignment of fibers (Figure 3(b)).

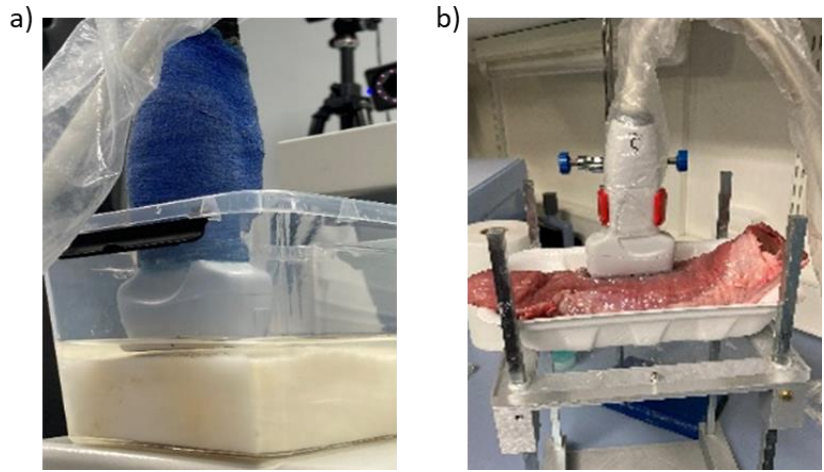

*Figure 3: a) SPB sequence tested on an isotropic PVA phantom. b) SPB sequence tested on ex vivo beef muscle. The transducer array was aligned parallel to the main axis of muscle fibers.*

*Mechanical testing*

To validate the quantification of the tensile anisotropy factor $\chi_E$ by using the novel SPB sequence, we conducted *ex vivo* experiments on 6 freshly excised porcine muscles (*iliopsoas*). These porcine muscles were excised from 3 healthy domestic pigs with an age of 3 to 4 months and a weight of 23 to 42 kg. In accordance with the national and European laws on animal protection, no approval for the study was needed, since the muscles were harvested after sacrifice in a pig involved for scientific purposes in another protocol approved by the local Ethical Committee and authorized by French authorities (APAFIS #22589-20190726115576) at the IHU-Strasbourg. Each *iliopsoas* muscle from the same pig was sampled into different pieces: 2 longitudinal (parallel) and 2 to 4 perpendicular (transverse) to the fibers. Each sample had a similar rectangular form (5x1 cm$^2$) (figure 4a). The fiber orientation was previously determined by classical B-mode ultrasound. Before stretching, both ultrasound sequences were applied to quantify mechanical parameters. The commercial sequence was used on both samples to quantify both shear moduli $\mu_{//SSI}$ and $\mu_{\perp SSI}$. The SPB sequence was used on the longitudinal sample to quantify shear modulus parallel $\mu_{//SPB}$ and the anisotropy factor $b = \mu_{\perp SPB}.\chi_E$. The standard deviation of the ultrasound anisotropy factor was calculated as follows:

$$\Delta\chi_E = \frac{|\Delta b.\mu_{\perp SSI} - b.\Delta\mu_{\perp SSI}|}{\mu_{\perp SSI}^2} \qquad (13)$$

Then samples muscles were fixed to the mechanical testing device (Instron 5944, Norwood, MA, USA) on the two ends by the system of jaws (figure 4b). Longitudinal samples were used to have muscle fibers aligned with the main axis of the traction device. The parallel Young's modulus ($E_{//}$) was calculated by using the Hooke's law from the slope of the experimental stress-strain curve obtained. Stepwise stretches from 0 to 12% nominal strain at 1% step were induced. At each 1% strain step, the tensile test was carried out. Complementary, transverse samples were used to quantify with the same protocol the Young' modulus perpendicular ($E_\perp$) to fibers. Then the elastic anisotropy factor $\chi_{EMECH}$ was calculated by the ratio of both Young's moduli.

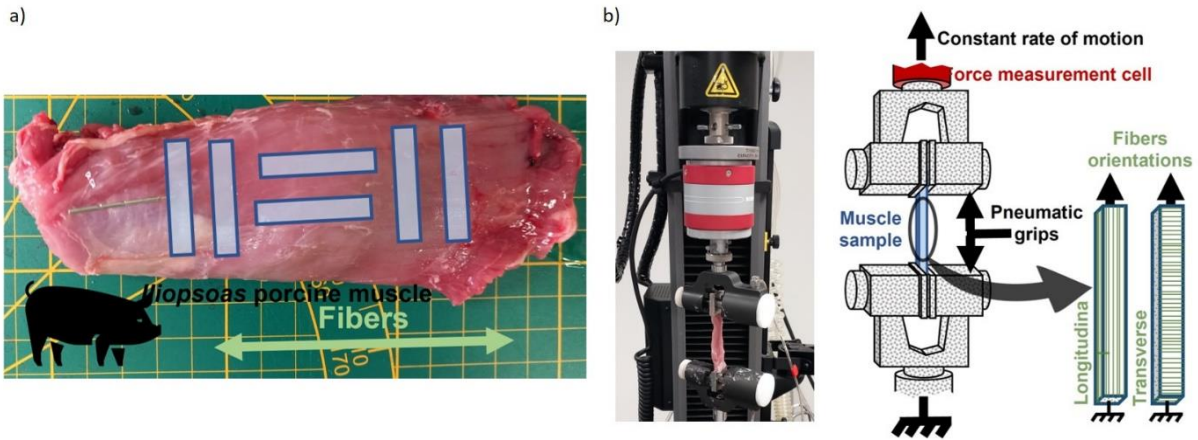

*Figure 4: a) 6 numerated samples extracted from one iliopsoas porcine muscle. b) Mechanical testing for one longitudinal muscle sample.*

*In vivo experiments*

Experiments were performed on 3 healthy volunteers (2 women, 1 man; age 28 ± 7 years; height 161 ± 12 cm; weight 56 ± 16 kg) with no self-reported history of upper limb neuromuscular and musculoskeletal disorders by following a protocol approved by the Ethics Committee of the University of Nantes (CERNI n°IRB: IORG0011023). Two muscle types were tested: a fusiform muscle, the *biceps brachii* (long head), and a pennated muscle, the *gastrocnemius medialis*. By using the B-mode image of the investigated muscle, the pinnation angle of each muscle was quantified between the main axis of fibers and the axis of the surface of the transducer array. This pinnation angle was used to fit with equation 9 the shear wave speed as a function of the angle of the steered pushing beam.

On the *biceps brachii*, scans were performed at a resting muscle length (approximately at slack angle), at 5% of the maximal isometric voluntary contraction level, and at a stretched muscle length. The SPB sequence allowed to quantify $\mu_{\perp SPBVIVO} \cdot \chi_{EVIVO}$ and $\mu_{//SPBVIVO}$ with the transducer array probe aligned along the muscle fibers, while the commercial sequence allowed the quantification of $\mu_{\perp SSIVIVO}$ with the probe aligned perpendicularly to the direction of muscle fibers and $\mu_{//SSIVIVO}$ with the probe aligned parallel to the fibers. These last measurements were only done in order to be used as a gold standard for the SPB sequence, to retrieve $\chi_{EVIVO}$ and to compare it with *ex vivo* results. Briefly, volunteers were seated in an isokinetic dynamometer (CMV AG, Dubendorf, Switzerland) chair and the participants were secured at the trunk with the shoulder abducted at 90° (Figure 5a). The forearm and wrist were positioned in pronation and neutral flexion-extension positions respectively. The medial epicondyle of the humerus was aligned with the axis of rotation of the dynamometer and used to estimate the elbow angle. The dynamometer was used to passively set the targeted elbow angle for measurements. The elbow joint was initially positioned at 90°, corresponding to a neutral flexion-extension position. Following an appropriate warm-up, participants were asked to perform two maximal voluntary isometric elbow flexions and extensions with the elbow in a neutral position (90°) with 1-min rest intervals between two trials. The maximal force registered across the elbow flexions was calculated and used to quantify submaximal isometric contraction force levels (percentages) of the maximal voluntary contraction (MVC). Both ultrasound sequences scans were performed on the mid-portion of the fusiform *biceps brachii* muscle (*i.e.,* long head) of three participants at three different muscle activation states: relaxed, stretching, and submaximal muscle contractions (5% and 10 % of MVC). For each state, the quantification of the mechanical parameters was performed with the probe aligned with the longitudinal direction of the muscle fibers and then rotated by 90° to be perpendicular to the muscle fibers. Custom-made probe holders were designed to minimize the compression during acquisitions and to facilitate the 90° rotation while avoiding motion in other space planes. For the submaximal contractions, participants were given visual feedback of the target force level and were asked to maintain it for approximately 10 s, corresponding to the required acquisition time for the elastography data. Force data was acquired by using a Biopac system (Biopac Systems Inc, Goleta, CA, USA). A visual feedback of the submaximal force levels was displayed in real-time on a screen to help the volunteer.

Because the *gastrocnemius medialis* is a pinnate muscle, only the SPB sequence was performed. Briefly, the participant seated in an isokinetic dynamometer (CMV AG, Dubendorf, Switzerland) chair with the hips flexed at 150° and the right knee fully extended. The chest and the waist were strapped to minimize trunk motion during acquisitions. The neutral position of the ankle was defined as an angle of 90° between the footplate and the shank. The lateral malleolus was used to estimate the center of rotation of the ankle and was aligned with the axis of the dynamometer. The foot was firmly strapped such that the potential heel displacement (Figure 5b). The probe was placed aligned with the direction of the muscle fibers of the *gastrocnemius medialis* and passively secured with a custom-made probe holder to minimize the pressure. As for the *biceps brachii*, measurements were then performed using the SPB method at a neutral plantar-dorsiflexion angle (rest) and during submaximal contractions at 5% and 10 % of MVC. Prior to measurements, contractions at MVC were performed to quantify the submaximal force levels used for visual feedback.

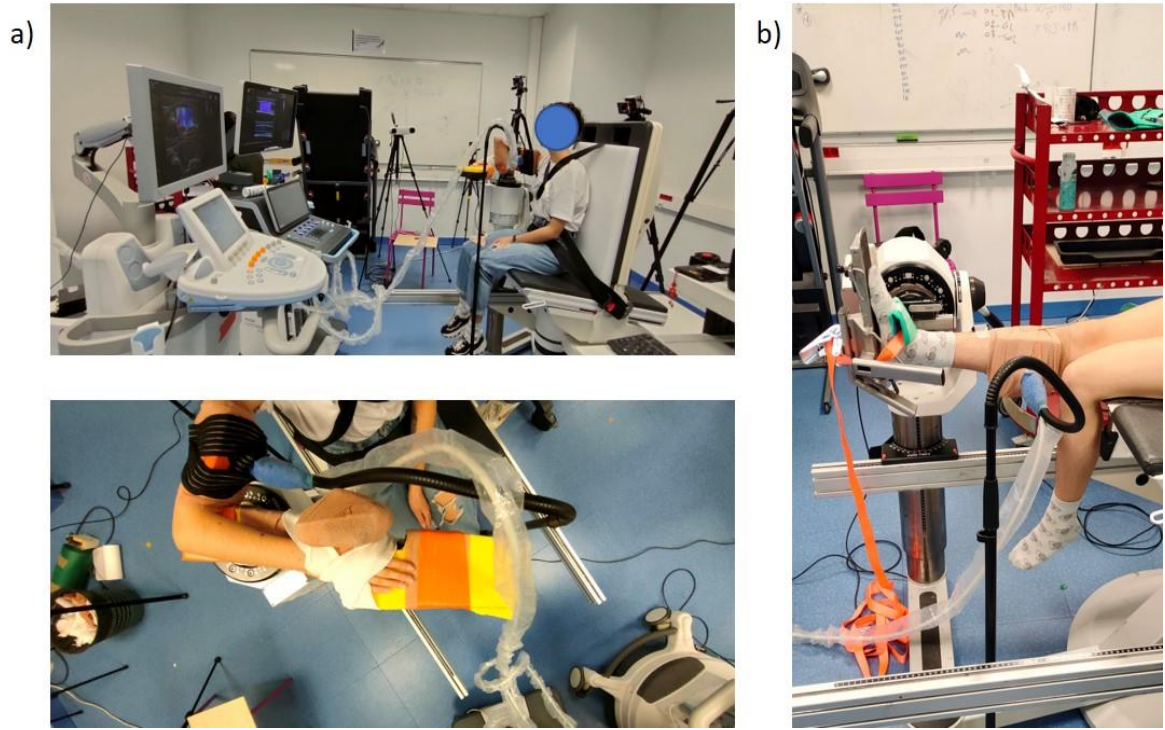

*Figure 5: a) A volunteer seated in an isokinetic dynamometer to investigate the anisotropy factor in a fusiform muscle, the biceps brachii. b) Another volunteer with the same device used for a pennated muscle the gastrocnemius medialis.*

**Results**

*Ex vivo and in vitro quantification of the anisotropy factor*

We firstly quantified the shear moduli parallel $\mu_{//SSI}$ and perpendicular $\mu_{\perp SSI}$ to fibers in *ex vivo* beef fusiform muscles using the commercial sequence of the Mach 30 device. The figure 6 shows these two acquisitions in one muscle. The mean shear modulus and associated standard deviation values were measured within the region of interest. The same measurements were performed similarly *in vivo* in the *biceps brachii*.

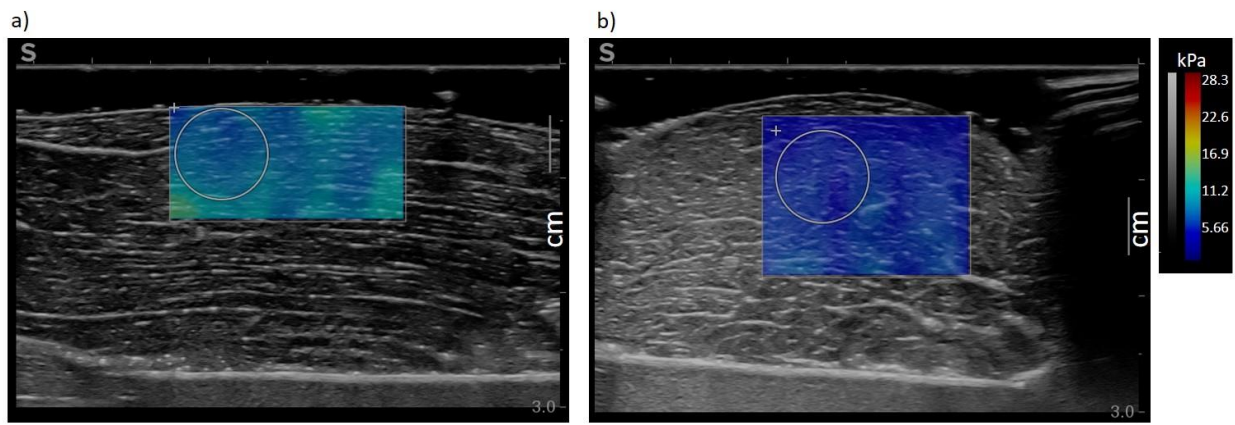

*Figure 6: Shear wave speed map in ex vivo fusiform beef muscle parallel a) and perpendicular b) to the fibers. The ROI were used to quantify the means values and their standard deviation. a) 2.76 ± 0.26 m/s and b) 2.09 ± 0.19 m/s leading to shear moduli of a) 7.61 ± 0.71 kPa and b) 4.36 ± 0.39 kPa. To acquire both images the transducer array was turned by 90° as described figure 1 (a).*

Second, the SPB sequence was applied in both the PVA gel phantom and a fusiform *ex vivo* beef muscle. As an example, the shear wave propagation is presented on figure 7 for 3 different steering angles (-20°, 0°, 20°) in a PVA phantom and at -10° in a fusiform *ex vivo* muscle as a function of time.

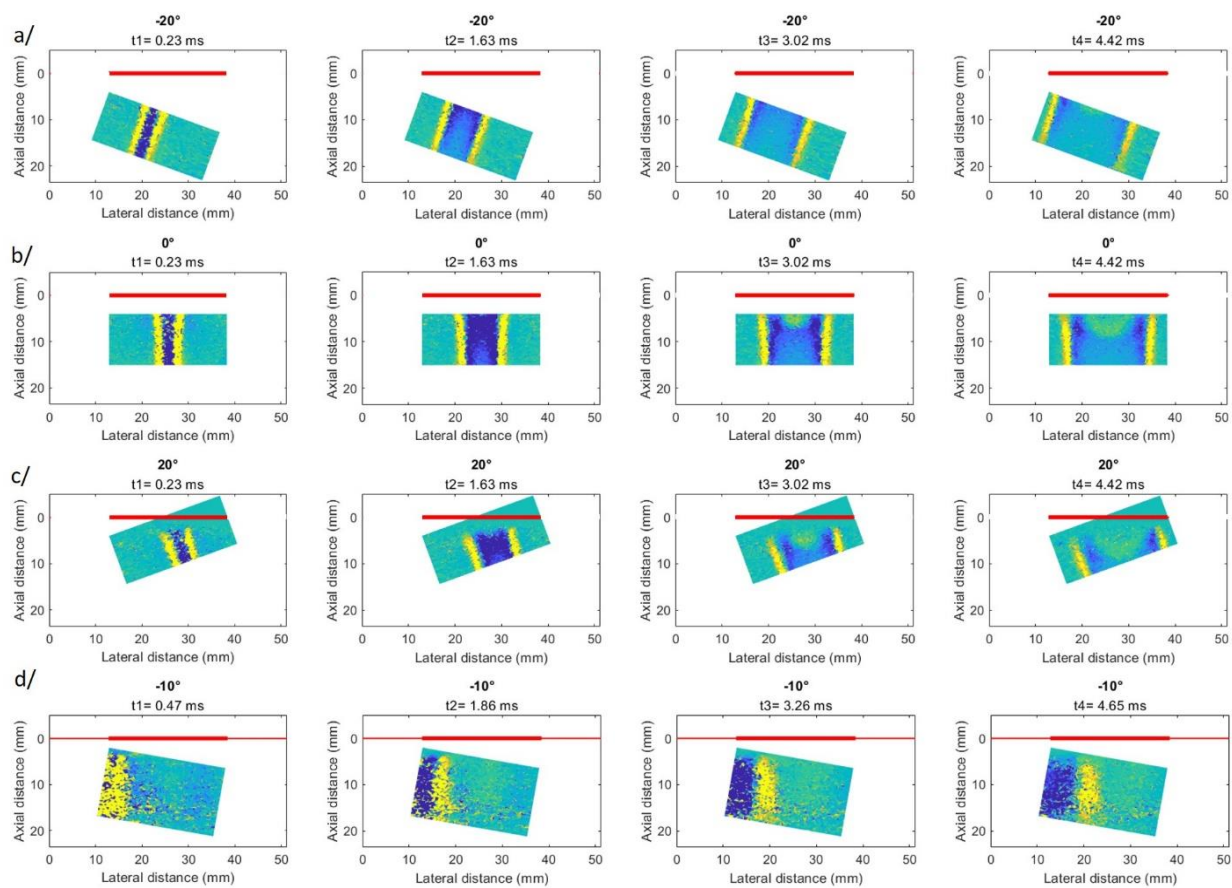

*Figure 7: Displacements field of the shear wave during propagation as a function of time for three different insonification angles of the steered pushing beams in a PVA phantom: a) -20°, b) 0°, c) 20° and at -10° in in vivo biceps brachii. The solid red line corresponds to the front head of the ultrasonic array.*

Then, the shear wave speed was quantified from the shear wave propagation movies as a function of the steering angle of the pushing beam, as presented in Figure 8 for both PVA phantom and *ex vivo* muscle. The results showed that, regardless of the steered push angle, the shear velocity is almost constant in the phantom (values of 8.09 ± 0.17 m/s) because of the isotropy of the medium (figure 8a). In the *ex vivo* muscle, a variation of the shear wave speed was observed as a function of the steering push beam angle (figure 8b), with values ranging from 3.48 ± 0.20 m/s to 6.22 ± 0.54 m/s. Consequently, these results showed that the SPB sequence is highly sensitive to the elastic anisotropy of the medium.

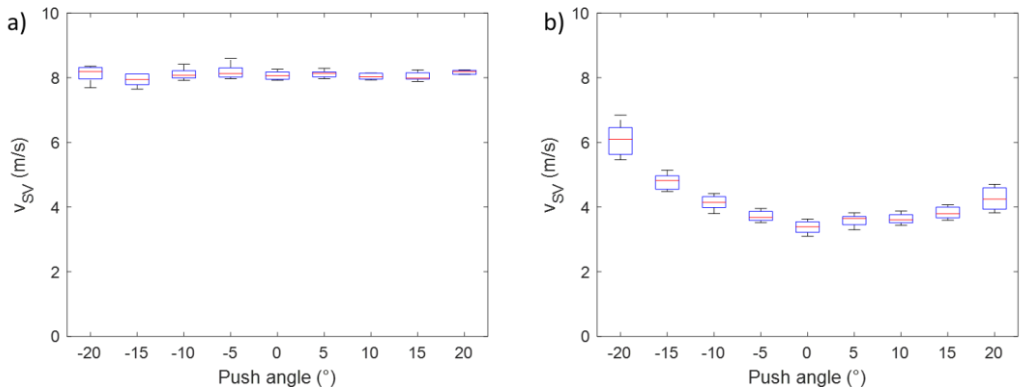

*Figure 8: Shear wave speed as a function of the angle of the steering push beam in isotropic PVA phantom a) and in ex vivo fusiform beef muscle b).*

*Ex vivo acquisitions coupled with tensile testing*

After being scanned with both previously-described ultrasound sequences, 6 *iliopsoas* fusiform muscles were cut and tested in a mechanical tensile device. At the exact locations where the samples are to be cut either parallel (longitudinal sample) and or perpendicular (transverse sample) to the fibers, B-mode images are extracted from the same muscle, as illustrated in figure 9a in a longitudinal and a transverse. Before cutting, the commercial sequence was used to quantify directly the shear modulus perpendicular to fibers $\mu_{\perp SSI}$. In the transverse samples. Similarly, before cutting, the SPB sequence was used to quantify the shear wave velocity as a function of the steering pushing beam angle in the longitudinal samples (figure 9b) as well as the modulus parallel to fibers $\mu_{//SSI}$ by using the commercial sequence. The red plots represent experimental data that are fit with equation 10 allowing to retrieve $\mu_{//SPB}$ and $\mu_{\perp SPB}.\chi_E$. The blue solid curve represents the fit equation 10 assuming that the angle of fibers (*i.e.* the fiber main axis is parallel to the surface of the transducer array) is set to 0°. Results over the 6 muscles are summarized in table 1.

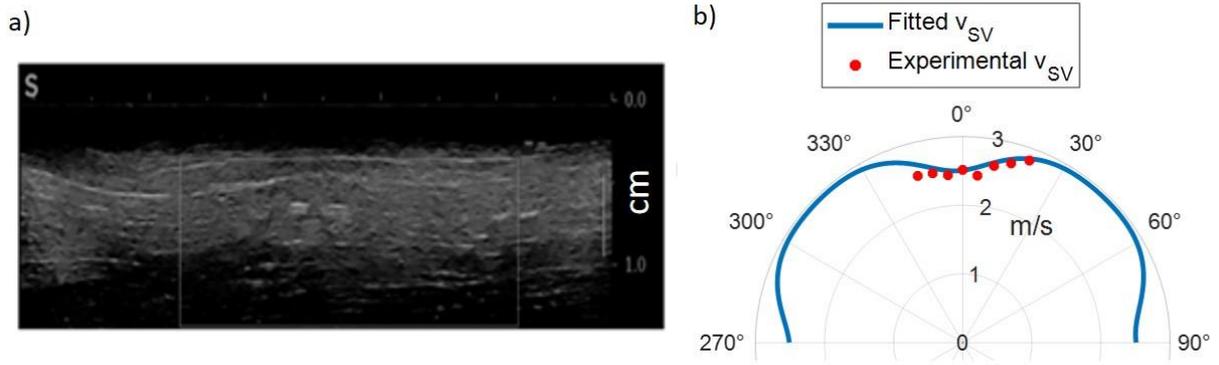

Figure 9: a) B-mode image extracted sample from one iliopsoas muscle along the fibers. b) Shear wave velocity data (red plot) fit by equation of the shear vertical mode (eq. 9, solid blue curve) as a function of the angle of the steered pushing beam.

In a second step, after cutting, all the samples were stretched using the mechanical device to quantify the Young's moduli either perpendicular or parallel to the fibers. The figure 10 illustrates the typical stress-strain curve obtained from two longitudinal and transverse samples from the same muscle. The mean Young's moduli and standard deviations were extracted from the linear fitting of these relationships at small strains, i.e. in the strain range where the material's mechanical behavior in tensile is linear and its elasticity is therefore independent of strain level (after fibers initial alignment but before nonlinear stiffening with strain, for tensile nominal strains between 2 and 8%). Results over the 6 muscles are summarized in table 1.

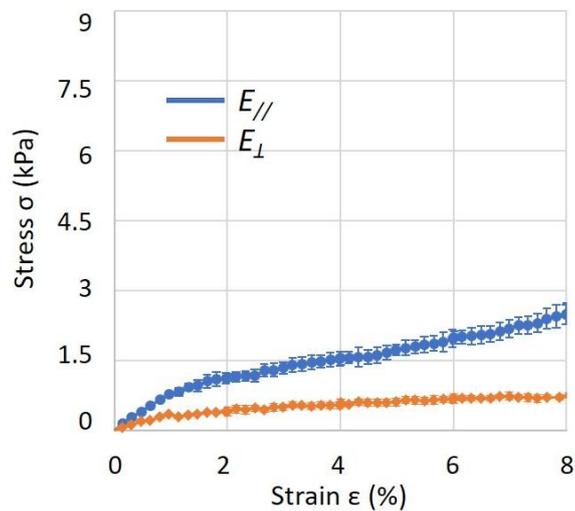

Figure 10: Stress-strain relationships obtained for two sample extracted from the same muscle perpendicular and parallel to the main fiber axis. The Young's moduli were calculated by fitting the stress/strain curve from 2 to 8%.

Table 1: Mechanical parameters in 6 ex vivo muscles determined with both ultrasound sequence and mechanical testing.

| Mechanical parameters (kPa) | | *Ex vivo* muscle 1 | *Ex vivo* muscle 2 | *Ex vivo* muscle 3 | *Ex vivo* muscle 4 | *Ex vivo* muscle 5 | *Ex vivo* muscle 6 |
|---|---|---|---|---|---|---|---|
| Ultrasound commercial sequence SSI | $\mu_{\perp SSI}$ | 9.33 ± 0.36 | 7.59 ± 0.29 | 3.61 ± 0.60 | 3.56 ± 0.46 | 4.26 ± 0.45 | 3.49 ± 0.66 |
| | $\mu_{//SSI}$ | 21.65 ± 0.08 | 8.00 ± 0.05 | 5.99 ± 0.82 | 6.72 ± 1.19 | 7.66 ± 0.64 | 6.88 ± 0.35 |

| Ultrasound research sequence SPB | $\mu_{//SPB}$ | 20.16 ± 0.74 | 9.11 ± 4.7 | 6.26 ± 0.27 | 7.15 ± 0.54 | 8.72 ± 0.55 | 6.65 ± 0.44 |
|---|---|---|---|---|---|---|---|
| | $\chi_E.\mu_{\perp SPB}$ | 28.48 ± 2.64 | 23.18 ± 1.97 | 8.62 ± 0.63 | 10.61 ± 1.17 | 11.79 ± 0.99 | 9.11 ± 0.78 |
| Tensile tests | $\chi_{EMECH}$ | 2.90 ± 0.43 | 2.93 ± 0.27 | 3.54 ± 1.08 | 3.44 ± 0.25 | 2.95 ± 0.34 | 2.55 ± 0.05 |
| Ultrasound anisotropy factor | $\dfrac{\chi_E.\mu_{\perp SPB}}{\mu_{\perp SSI}}$ | 3.05 ± 0.17 | 3.05 ± 0.14 | 2.38 ± 0.22 | 3.00 ± 0.05 | 2.77 ± 0.06 | 2.61 ± 0.27 |

Table 1 shows a comparison between shear moduli measured with the commercial sequence (SSI) and with the SPB sequence. A Mann-Whitney test was performed (p=0.818) showing no significant difference between SSI and SPB measurements. Table 1 illustrates the elastic anisotropic ratio $\chi_E$ measured from mechanical testing and the elastic anisotropic ratio calculated from the parameters extracted from the SPB sequence ($\chi_E.\mu_{\perp SPB}$) and divided by the shear modulus perpendicular to fiber measured with the commercial sequence ($\mu_{\perp SSI}$). Once again, a Mann-Whitney test was performed (p=0.558) showing no significant difference between both results.

*In vivo experiments*

*In vivo* scans were performed on both *biceps brachii* (fusiform muscle) and the *gastrocnemius medialis* (pennated muscle). The figure 11 shows results for 1 volunteer at 2 levels of contraction, at rest and at 10% MVC, and during stretching. On the B-mode image (figure 11(a, b, c)), the main axis of fibers was determined in order to fit the shear wave speed data with equation 10. On figure 11(d, e, f) the experimental shear wave velocity data are plotted in red and the corresponding fitting with equation 9 drawn in solid blue line. The whole results on *biceps brachii* are presented in table 2. A Mann-Whitney test was performed (p=0.977) revealed no significant differences between both SSI and SPB measurements.

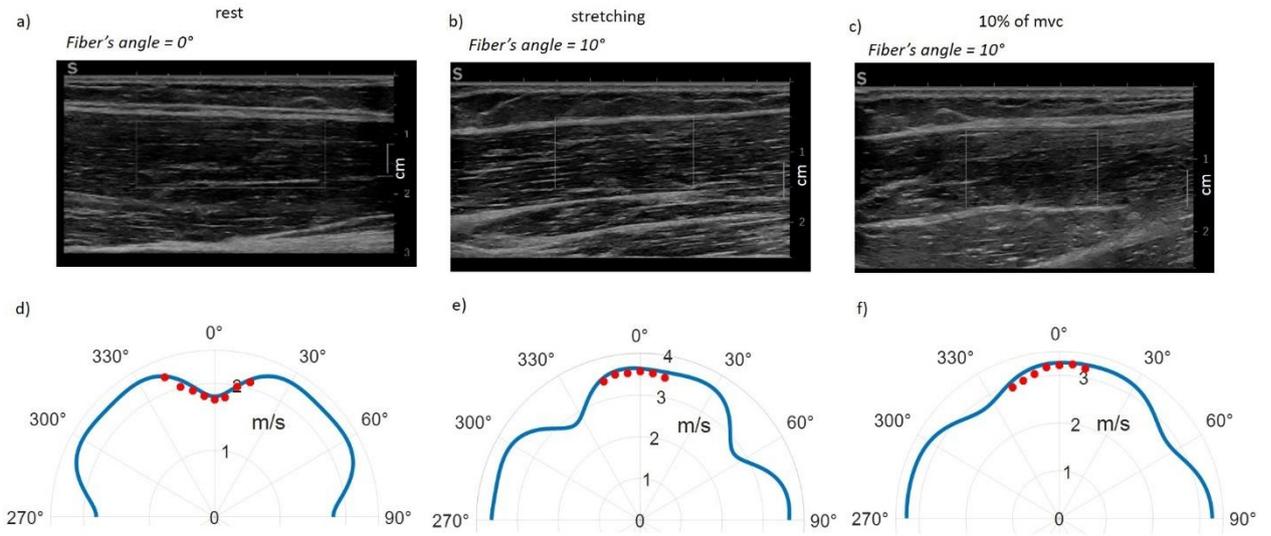

*Figure 11: a, b, c) B-mode image of the biceps brachii for one volunteer at rest, during stretching and at 10% of MVC. d, e, f) Shear wave velocity data in red are fit by equation 9 in blue.*

*Table 2: Mechanical parameters determined with both ultrasound sequences in 3 healthy volunteers for different level of biceps brachii contraction.*

| Volunteer # | % MVC | $\mu_{\perp SSIVIVO}$ (kPa) | $\mu_{//SSIVIVO}$ (kPa) | $\mu_{//SPBVIVO}$ (kPa) | $\chi_{EVIVO} \cdot \mu_{\perp SPBVIVO}$ | $\dfrac{\chi_{EVIVO} \cdot \mu_{\perp PSBVIVO}}{\mu_{\perp SSIVIVO}}$ |
|---|---|---|---|---|---|---|
| #1 | Stretching | 1.94 ± 0.23 | 11.6 ± 1.06 | 10.19 ± 0.38 | 6.07 ± 0.89 | 3.13 ± 0.08 |
| | 0 | 1.75 ± 0.20 | 3.24 ± 0.28 | 2.80 ± 0.21 | 4.54 ± 0.51 | 2.56 ± 0.24 |
| | 5 | 2.01 ± 0.14 | 8.47 ± 0.73 | 10.18 ± 0.36 | 7.01 ± 1.62 | 2.90 ± 0.17 |
| | 10 | 4.88 ± 1.20 | 21.36 ± 2.65 | 21.86 ± 1.04 | 15.21 ± 4.25 | 3.12 ± 0.10 |
| #2 | Stretching | 2.70 ± 0.61 | 13.88 ± 1.59 | 12.95 ± 0.21 | 7.74 ± 0.43 | 2.86 ± 0.49 |
| | 0 | 2.38 ± 0.72 | 3.62 ± 0.42 | 3.27 ± 0.14 | 5.31 ± 0.32 | 2.23 ± 0.54 |
| | 5 | 1.60 ± 0.27 | 4.16 ± 0.36 | 3.03 ± 0.08 | 4.04 ± 0.19 | 2.92 ± 0.23 |
| | 10 | 2.92 ± 0.97 | 7.63 ± 1.52 | 10.69 ± 0.23 | 7.82 ± 1.53 | 2.67 ± 0.36 |
| #3 | Stretching | 2.02 ± 0.44 | 9.87 ± 1.31 | 9.06 ± 0.23 | 6.21 ± 1.04 | 3.08 ± 0.16 |
| | 0 | 1.85 ± 0.24 | 3.82 ± 0.60 | 3.02 ± 0.12 | 4.94 ± 0.38 | 2.67 ± 0.14 |
| | 5 | 2.41 ± 0.43 | 4.41 ± 0.63 | 4.48 ± 0.12 | 5.65 ± 0.28 | 2.34 ± 0.30 |
| | 10 | 2.83 ± 0.37 | 6.2 ± 0.63 | 9.33 ± 0.61 | 6.22 ± 2.30 | 2.19 ± 0.52 |

The figure 12 illustrates a typical example for a pennated muscle, the *gastrocnemius medialis* at rest and at 2 contraction levels (5 and at 10% of MVC). On the B-mode image (figure 12(a, b, c)), the main axis of the fibers was determined in order to fit the shear wave speed data with equation 10. On figure 12(d, e, f) the experimental shear wave velocity data are plotted in red and the corresponding fitting with equation 10 drawn in solid blue line. One can notice that the angle of fibers changes significantly with contraction, with values ranging from 22° to 33°. The whole results on *gastrocnemius medialis* are presented in table 3.

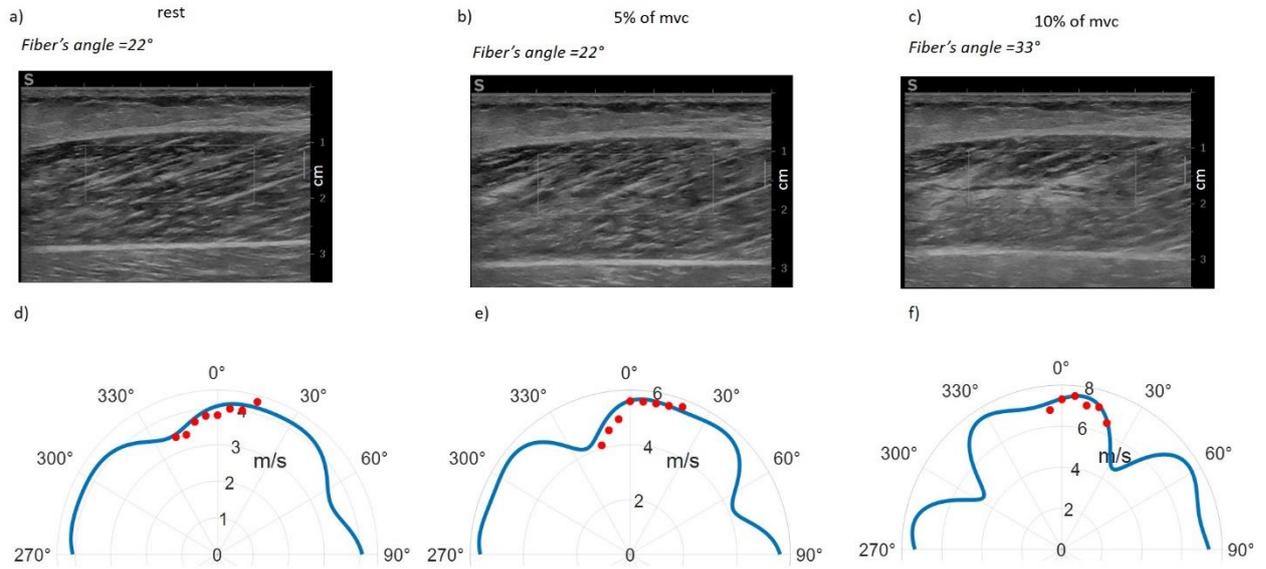

*Figure 12: a, b, c) B-mode image of the gastrocnemius medialis for one volunteer at rest and at 5% and 10% of MVC showing 22°, 22° and 33° of pinnation angle regarding the surface of the transducer array. d, e, f) Shear wave velocity data in red and fit by equation 12 in blue. The fitting was adjusted by the pinnation angle determined on the B-mode image.*

*Table 3: Mechanical parameters determined with both ultrasound sequences for one healthy volunteer for different level of gastrocnemius medialis contraction.*

| Volunteer # | % MVC | $\mu_{//SPBVIVO}$ (kPa) | $\chi_{EVIVO} \cdot \mu_{\perp SPBVIVO}$ |
|---|---|---|---|
| # 1 | 0 | 17.28 ± 0.68 | 12.01 ± 1.17 |
|  | 5 | 31.01 ± 2.27 | 17.54 ± 3.66 |
|  | 10 | 21.78 ± 5.96 | 49.54 ± 17.39 |
| # 2 | 0 | 16.71 ± 0.36 | 7.02 ± 0.21 |
|  | 5 | 18.28 ± 5.00 | 33.87 ± 11.55 |
|  | 10 | 21.11 ± 3.94 | 33.75 ± 7.82 |
| # 3 | 0 | 16.26 ± 0.66 | 9.77 ± 0.99 |
|  | 5 | 28.35 ± 1.27 | 17.51 ± 2.03 |
|  | 10 | 34.10 ± 1.74 | 20.07 ± 2.89 |

**Discussion**

The validity of the novel SPB sequence was firstly demonstrated on phantoms. On an isotropic phantom, the angle of the steered push beam was changed from -20° to 20°, with a 5° increment. Results showed that the shear wave velocity was not impacted with the angle of the pushing beam. Then, the SPB sequence was tested in an *ex vivo* fusiform beef muscle with the ultrasound transducer array parallel to the main axis of the fibers muscle. Results presented a change in the shear wave velocity value, showing a clear influence of the transverse isotropy of the medium in shear elastic parameters as predicted by the theoretical framework (eq. 12). By using eq. 10, the shear modulus parallel to fibers $\mu_{//SPB}$ as well as the anisotropy factor $\mu_{\perp SPB}.\chi_E$ was estimated. By comparison of $\mu_{//SPB}$ with the shear modulus $\mu_{//SSI}$ measured with the commercial sequence, no significant difference was observed (p=0.818). To confirm these results in terms of anisotropy factor, mechanical testing was performed on 6 *ex vivo* beef muscle samples to quantify the young's moduli ratio $\chi_{EMECH}$. By dividing the anisotropy factor by the shear modulus perpendicular to fibers $\mu_{\perp SSI}$ obtained with the commercial sequence, a good accordance was observed between $\chi_{EMECH}$ and $\chi_E$ from the tensile tests and SPB method respectively. These results showed that it is possible to use the new SPB sequence with the transducer array parallel to the main fiber axis to quantify not only the shear modulus parallel to fibers, but also the elastic anisotropy factor. This last innovative parameter can then be used as a new potential biomarker of muscle structural properties and has the potential to serve as a biomarker of muscle pathology. This is a strong advantage compared to other existing techniques, which involve either rotation of the transducer array and multiple acquisitions (Gennisson *et al.*, 2010; Lee *et al.*, 2012) or need to have access to complex devices for a volumetric acquisition (Wang *et al.*, 2013; Gennisson *et al.*, 2015) to quantify an anisotropy factor with a standard ultrasound elastography device. More importantly, these results open the possibility to quantify $\mu_{//SPB}$ and $\mu_{\perp SPB}.\chi_E$ in pennated muscles (i.e., the vast majority of muscles).

In a second step, *in vivo* experiments were performed on volunteers on 2 types of muscle: a fusiform muscle (*biceps braachi*) and a pennated muscle (*gastrocnemius medialis*). The results show that it was possible to quantify the anisotropy factor in both muscles during different states of contraction and stretching. For the fusiform muscle, it was possible to use the commercial sequence to evaluate the shear moduli both parallel and perpendicular to fibers. Similarly, to the *ex vivo* mechanical tests, it was then possible to quantify the Young' moduli ratio with the SPB sequence. *In vivo* results show a good accordance between the *ex vivo* ($\chi_E$ from 2.55 ± 0.05 to 3.54 ± 1.08) and the *in vivo* ($\chi_{EVIVO}$ from 2.23 ± 0.54 to 3.13 ± 0.08) Young's moduli ratio. The results are fully coherent with those related by others studies on the same muscles. With magnetic resonance elastography, Guo *et al.* reported a Young's moduli ratio of 2.53 on 10 volunteers at rest in the *gastrocnemius medialis* (Guo *et al.*, 2016). On another muscle *in vivo* at rest, *vastus lateralis*, Knight *et al.* by using a volumetric acquisition ultrasound device, estimated the ratio $(E_{//} - E_\perp)/E_\perp = \chi_E - 1$. They found a value of 4.57 which is higher than the values reported here (Knight *et al.*, 2022). It can be explained by the physical architecture of each muscle. Indeed, the *biceps braachi* is a fusiform and *vastus lateralis* a pennated muscle, likely to have different mechanical properties and anisotropy coefficients. This emphasize the fact that the SPB technique is applicable only in fusiform or mono-pennated muscle. The quantification of the shear modulus parallel to fibers and of the anisotropy factor can only be done when the probe is aligned with the main axis of the muscle. This is a true drawback compared to 3D acquisition, nevertheless it allows to do the real-time acquisition with an ultrafast ultrasound device and to acquire these two biomarkers live during contraction or during controlled movements. However, it shows that the SPB technique is reliable to quantify anisotropy factor *in vivo*.

Moreover, results during different states of contraction (at rest, 5% and 10% of MVC) in both muscle types have been observed (tables 2 and 3). The first thing to notice from the SPB *in vivo* results is that not only the shear modulus but also the anisotropy factor increases with contraction. This is in line with previous works that used different approaches to assess this ratio (Shinohara et al., 2010; Gennisson et al., 2010). $\mu_{//SPB}$, and $\chi_E \cdot \mu_{\perp SPB}$ change for each depends on the contraction state as well as on the orientation of the muscle fibers. By using the fitting curve from eq. 10, it is then possible to retrieve the angle of the fibers directly by the quantification of the shear wave speed. It was already proposed by Lee et al., in cardiac muscle (Lee et al., 2012). But this is possible only when the transducer array is aligned with the main axis of the muscle, i.e. fusiform muscles or pennated muscle with a pinnation of the fiber within the imaging plane. However, this application has the great advantage to allow for a direct access to the fiber orientation with an unique acquisition technique without other dedicated algorithm or set of data.

Even if the SPB sequence presents some great advantage to investigate non-invasively and in real time and using a classical probe the anisotropic elastic parameters of the muscle, this approach has some limitations. The depth of investigation is limited with this linear transducer array (around 20 mm) due to the insonification angle that are restricted to a range from -20° to 20°. Implementation on other types of transducer arrays such as curved or phased arrays could make it possible to measure deeper in the organs and this option has to be investigated in the future. Then, it is also assumed that the shear modulus measured *in vivo* is homogeneous in space, which is a strong hypothesis assumed as a first approach. Last but not least, the use of this technique is not applicable to all sort of muscles, as previously mentioned, since the placement of the probe regarding the main fiber axis is quite important. As an example, in a bipinnate muscle such as *tibialis posterior*, for now only volumetric devices could acquire these parameters.

At last *in vivo* acquisitions on pennated muscle show an increase of the shear modulus parallel to fibers with the MVC as well as an increase of the anisotropic factor for all volunteers. These results are in good accordance with other studies (Gennisson *et al.*, 2010; Hug *et al.*, 2015) and show the potential of the technique to study these two parameters in real time during movements.

**Conclusion**

A new elastography ultrasound sequence was proposed to better characterize mechanical and structural properties of transverse isotropic skeletal muscles. The technique shows that it is possible in fusiform muscles or pennated muscles, when the transducer array is aligned with the main axis of fibers, to quantify the shear modulus parallel to fibers as well as an anisotropy factor (ratio of Young's moduli times shear modulus perpendicular to fibers) using a single and fast acquisition. In future works this approach will allow to investigate a new biomarker of the anisotropy during movement without the constraint of moving the transducer between acquisitions.

**Aknowledgement**

This work was funded by: ANR INNOVAN (ANR-19-CE19-0017), PEPS CNRS MUSCLOR 2022, PEDECIBA and ANII (FMV_1_2019_1_155527), France Life Imaging (ANR-11-INBS-0006 grant from the French "Investissements d'Avenir" program, and supported by IdEx Unistra (ANR-10-IDEX-0002) with the IRIS technology platform and by SFRI (STRAT'US project, ANR-20-SFRI-0012) under the framework of the French Investments for the Future Program. The authors also thank the team of the preclinical department of the IHU Strasbourg for their logistic support.

**References**

Bercoff J, Tanter M and Fink M 2004 Supersonic shear imaging: a new technique for soft tissue elasticity mapping *IEEE Trans. Ultrason. Ferroelectr. Freq. Control* **51** 396-409

Berg W, Cosgrove D, Doré C, Schäfer F, Svensson W, Hooley R, Ohlinger R, Mendelson E, Balu-Maestro C, Locatelli M, Tourasse C, Cavanaugh B, Juhan V, Stavros A, Tardivon A, Gay J, Henry J-P and Cohen-Bacrie C 2012 Shear-wave elastography improves the specificity of breast US: The BE1 Multinational Study of 939 Masses *Radiology* **262** 435-449

Caenen, A., Knight, A. E., Rouze, N. C., Bottenus, N. B., Segers, P., and Nightingale, K.R. 2020 Analysis of multiple shear wave modes in a nonlinear soft solid: Experiments and finite element simulations with a tilted acoustic radiation force. J. Mech. Behav. Biomed. Mat. **107**, 103754.

Catheline S, Wu F and Fink M 1999 A solution to diffraction biases in sonoelasticity: The acoustic impulse technique *J. Acoust. Soc. Am.* **105** 2941-2950

Chatelin S, Bernal M, Deffieux T, Papadacci C, Flaud P, Nahas A, Boccara C, Gennisson J-L, Tanter M and Pernot M 2014 Anisotropic polyvinyl alcohol hydrogel phantom for shear wave elastography in fibrous biological soft tissue: a multimodality characterization *Phys. Med. Bio.* **59** 6923–6940

Correia M, Deffieux T, Chatelin S, Provost J, Tanter M, and Pernot, M 2018 3D elastic tensor imaging in weakly transversely isotropic soft tissues *Phys. Med. Biol.* **63** 155005

Deffieux T, Montaldo G, Tanter M, and Fink M 2009 Shear wave spectroscopy for in vivo quantification of human soft tissues visco-elasticity *IEEE Trans Med Imaging* **28** 313-322

Deffieux T, Gennisson J-L, Larrat B, Fink M and Tanter M 2012 The variance of quantitative estimates in shear wave imaging: Theory and experiments *IEEE Trans. Ultrason. Ferroelectr. Freq. Control* **59** 2390-2410

Fatemi M and Greenleaf J 1998 Ultrasound-stimulated vibro-acoustic spectrography *Science* **280(5360)** 82-85

Fromageau J, Gennisson J-L, Schmitt C, Maurice RL, Mongrain R and Cloutier G 2007 Estimation of polyvinyl alcohol cryogel mechanical properties with four ultrasound elastography methods and comparison with gold standard testings *IEEE Trans. Ultrason. Ferroelectr. Freq. Control* **54** 498-509

Gennisson J-L, Deffieux T, Macé E, Montaldo G, Fink M and Tanter M 2010 Viscoelastic and anisotropic mechanical properties of in vivo muscle tissue assessed by supersonic shear imaging *Ultrasound Med. Biol.* **36** 789-801

Gennisson J-L, Provost J, Deffieux T, Papadacci C, Imbault M, Pernot M and Tanter M 2015 4-D ultrafast shear-wave imaging *IEEE Trans. Ultrason. Ferroelectr. Freq. Control* **62** 1059-1065

Grasland-Mongrain P, Mari J-M, Chapelon J-Y and Lafon C 2013 Lorentz force electrical impedance tomography *Innovation and Research in Biomedical Engineering* **34** 357-360

Guo J, Hirsch S, Scheel M, Braun J and Sack I 2016 Three-Parameter Shear Wave Inversion in MR Elastography of Incompressible Transverse Isotropic Media: Application to In Vivo Lower Leg Muscles *Magnetic Resonance in Medicine* **75** 1537–1545

Hug F, Tucker K, Gennisson J-L, Tanter M and Nordez A 2015 Elastography for Muscle Biomechanics: Toward the Estimation of Individual Muscle Force *Exercise and Sport Sciences Review* **43** 125-133

Knight A, Trutna C, Rouze N, Hobson-Webb L, Caenen A, Jin F, Palmeri M and Nightingale K 2022 Full Characterization of in vivo Muscle as an Elastic, Incompressible, Transversely Isotropic Material using Ultrasonic Rotational 3D Shear Wave Elasticity Imaging *IEEE Trans. Med. Imaging* **41** 133–144.

Lee W-N, Larrat B, Pernot M and Tanter M 2012 Ultrasound elastic tensor imaging: Comparison with MR diffusion tensor imaging in the myocardium *Phys. Med. Biol*. **57** 5075–5095

Loupas T, Powers JT and Gill RW 1995 An axial velocity estimator for ultrasound blood flow imaging, based on a full evaluation of the Doppler equation by means of a two-dimensional autocorrelation approach *IEEE Trans. Ultrason. Ferroelectr. Freq. Control* **42** 672-688

Montaldo G, Tanter M, Bercoff J, Benech N and Fink M 2009 Coherent plane-wave compounding for very high frame rate ultrasonography and transient elastography *IEEE Trans. Ultrason. Ferroelectr. Freq. Control* **56** 489–506

Muthupillai R, Lomas DJ, Rossman PJ, Greenleaf JF, Manduca A and Ehman RL 1995 Magnetic resonance elastography by direct visualization of propagating acoustic strain waves *Science* **269**(5232) 1854–1857.

Ngo H.-H.-P., Poulard T., Brum J. and Gennisson J.L. 2022 Anisotropy in ultrasound shear wave elastography: An add-on to muscles characterization *Frontiers in Physics* **13** 1000612

Nightingale K, McAleavey S and Trahey G 2003 Shear-wave generation using acoustic radiation force: in vivo and ex vivo results *Ultrasound Med. Biol.* **29** 1715-23

Parker K, Huang S, Musulin R and Lerner R 1990 Tissue response to mechanical vibrations for "sonoelasticity imaging" *Ultrasound Med. Biol*. **16** 241-246

Rouze, N. C., Wang, M. H., Palmeri, M. L., and Nightingale, K. R. 2013 Finite element modeling of impulsive excitation and shear wave propagation in an incompressible, transversely isotropic medium. J. Biomech. **46**, 2761–2768.

Rouze, N. C., Palmeri, M. L., and Nightingale, K. R. 2020 Tractable calculation of the Green's tensor for shear wave propagation in an incompressible, transversely isotropic material. Phys. Med. Biol. **65**, 015014

Royer D and Dieulesaint E 1999 Elastic Waves in Solids. I. Free and Guided Propagation (Berlin: Springer) Chap. 4

Royer D, Gennisson J-L, Deffieux T and Tanter M 2011 On the elasticity of transverse isotropic soft tissues (L) *J. Acoust. Soc. Am.* **129** 2757-2760

Sandrin L, Tanter M, Gennisso J-L, Catheline S and Fink M 2002 Shear elasticity probe for soft tissue with 1D transient elastography *IEEE Trans. Ultrason. Ferroelectr. Freq. Control* **49** 436-446

Sandrin L, Fourquet B, Hasquenoph J-M, Yon S, Fournier C, Mal F, Christidis C, Ziol M, Poulet B, Kazemi F, Beaugrand M and Palau R 2003 Transient elastography: A new noninvasive method for assessment of hepatic fibrosis *Ultrasound Med. Biol.* **29** 1705-1713

Sarvazyan A, Skovoroda A, Emelianov S, Fowlkes J, Pipe J, Adler R, Buxton R and Carson P 1995 Biophysical bases of elasticity imaging *Acoustical imaging* (Boston, MA: Springer) 223-240

Shinohara M, Sabra K, Gennisson J-L, Fink M and Tanter M 2010 Real-time visualization of muscle stiffness distribution with ultrasound shear wave imaging during muscle contraction *Muscle Nerve* **42** 438-441

Song P, Zhao H, Manduca A, Urban M, Greenleaf J and Chen S 2012 Comb-push ultrasound shear elastography (CUSE): a novel method for two-dimensional shear elasticity imaging of soft tissues *IEEE Trans. Med. Imaging* **31** 1821-1832

Thomsen L 1986 Weak elastic anisotropy *Geophysics* **51** 1954-1966

Urban M, Chalek C, Kinnick R, Kinter T, Haider B, Greenleaf J, Thomenius K and Fatemi M 2011 Implementation of vibro-acoustography on a clinical ultrasound system *IEEE Trans. Ultrason. Ferroelectr. Freq. Control* **58** 1169-1181

Wang M, Byram B, Palmeri M, Rouze N and Nightingale K 2013 Imaging transverse isotropic properties of muscle by monitoring acoustic radiation force induced shear waves using a 2-D matrix ultrasound array. *IEEE Trans. Med. Imaging* **32** 1671-1684